\documentclass[a4paper, 12pt]{article}
\usepackage[a4paper,margin=3.5cm]{geometry}
\usepackage[english]{babel}
\usepackage{graphicx}
\usepackage{amsmath}
\usepackage{color}
\usepackage[usenames,dvipsnames]{xcolor}
\usepackage{amssymb}
\usepackage{amsthm}
\usepackage{morefloats}
\usepackage{caption}
\usepackage{fancyhdr}
\usepackage{footnote}
\usepackage{subfig}
\usepackage{multicol}
\usepackage{hyperref}
\newcommand\Unif{\textrm{Unif}}
\newcommand\Gauss{\textrm{Gauss}}

\begin{document}

\title{BlurRing}

\author{Lydia Brenner and Carsten Burgard}
\date{\today}
\maketitle
\thispagestyle{empty}

\begin{abstract}
A code package, BlurRing, is developed as a method to allow for
multi-dimensional likelihood visualisation. From the BlurRing
visualisation additional information about the likelihood can be
extracted. The spread in any direction of the overlaid likelihood
curves gives information about the uncertainty on the confidence
intervals presented in the two-dimensional likelihood plots.
\end{abstract}

\begin{multicols}{2}

\section*{Introduction}
\noindent The predictive power of the Standard Model (SM) of particle
physics has, so far, been confirmed with every measurement at the
Large Hadron Collider (LHC). With the discovery of the Higgs boson in
2012 by the ATLAS and CMS collaboration \cite{Aad:2012tfa,Chatrchyan:2012xdj}, the SM is able to map
theoretical prediction to experimental measurements with a level of
precision that has not been seen before. Due to the discovery of the
Higgs boson, new ways to look at the data coming from the LHC have
become of high interest to the experiments. However many tools are not
equipped to handle these new methods. One of the main complications
with the new type of measurements performed at the LHC comes from the
inability to represent multi-dimensional likelihoods. While previously
measurements, and in particular searches for new particles, had one or
two parameters of interest, current measurements take more than two
parameters into account.

While it is perfectly possible to represent likelihoods in one or two
dimensions without approximating the models, it is not possible to
represent likelihoods with higher dimensionality in a comprehensive
way on a two dimensional paper. The current method for
multi-dimensional likelihoods profiles the remaining parameters
\cite{cowan1998statistical}. This method fits all parameters simultaneously and uses the best
fitted value for the remaining, i.e.~not plotted, parameters. The
profiling method incorporates the remaining parameters into the
likelihood, rather than fixing the remaining parameters to their
expected values. However, it does not show how the likelihood changes
due to possible changes in the values of the other parameters.

Furthermore, the profiled likelihood curve in two dimensions does not
necessarily map to a continuous path in the full parameter space.  To
incorporate this missing information in plotting higher dimensional
likelihoods, the BlurRing method is developed.  The BlurRing method is
explained in detail below, and the full package available through
\href{https://tinyurl.com/BlrRng}{tinyurl.com/BlrRng} and fully  implemented in RooFit \cite{roofit}.

\section*{Method}
\begin{figure*}[!ht]
\noindent \begin{centering}
\includegraphics[width=.45\textwidth]{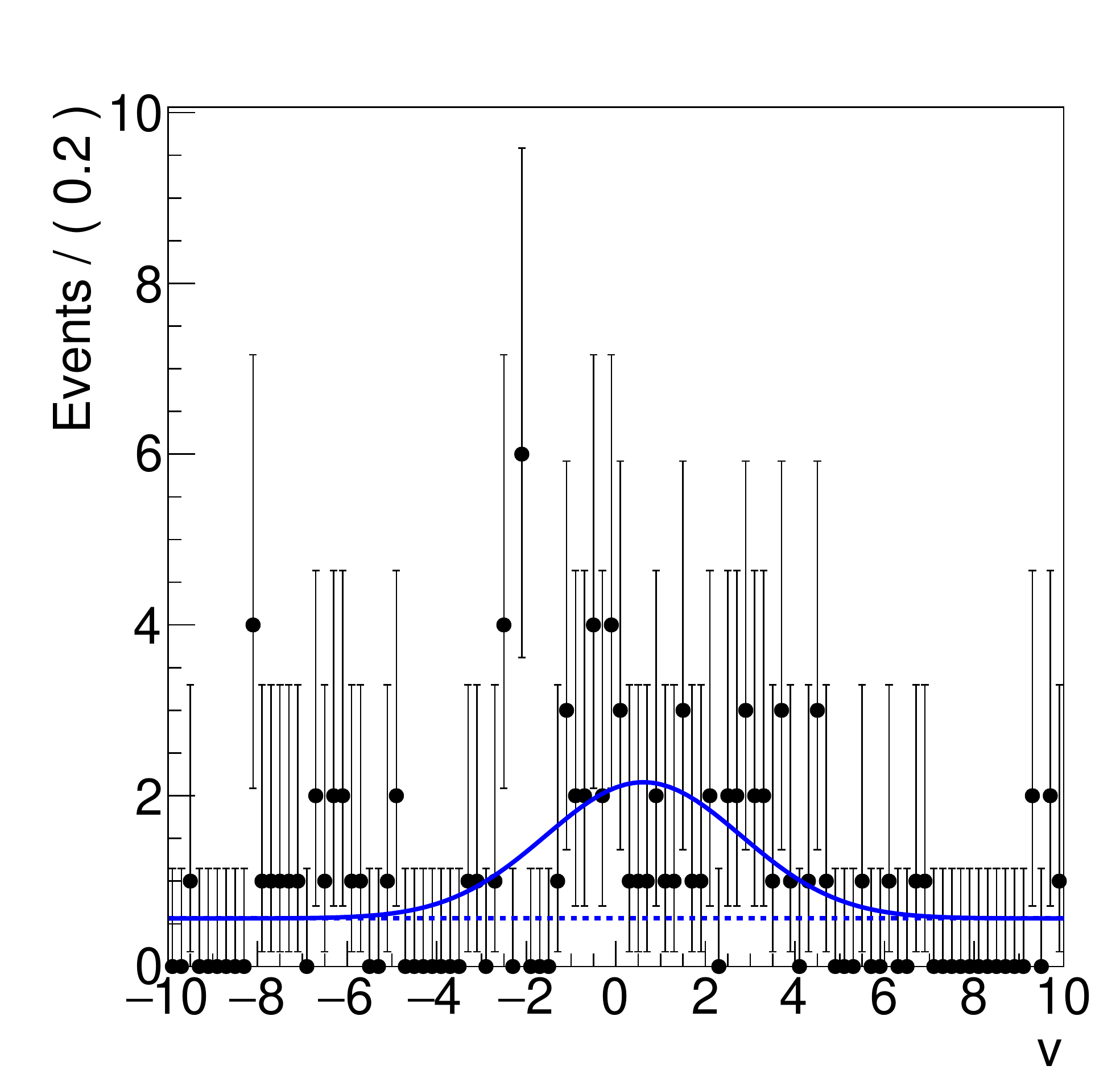}
 \includegraphics[width=.45\textwidth]{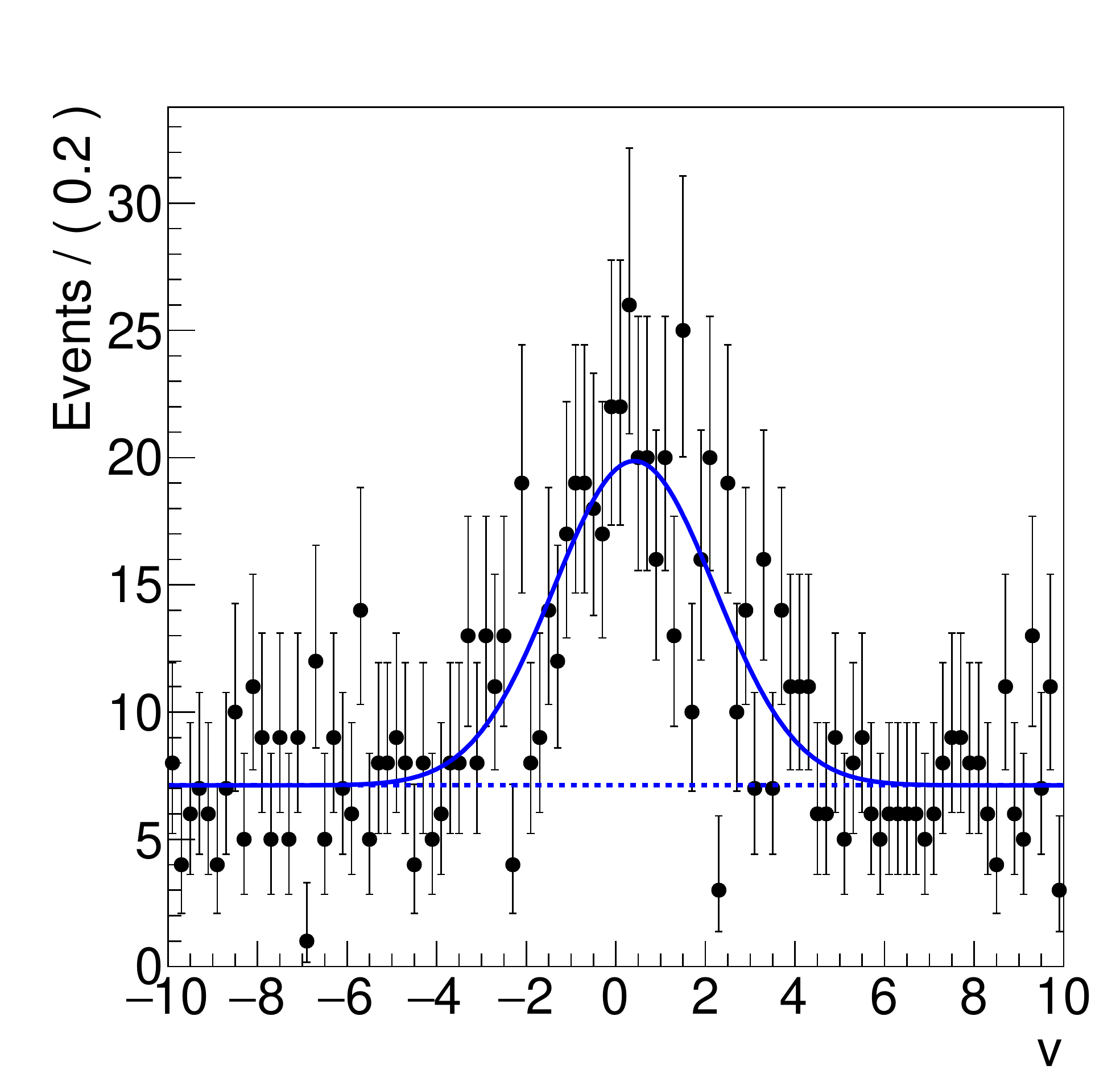}
 \par\end{centering}
\caption{Example model with uniform background and Gaussian signal for 100 events (left) and 1000 events (right).}
\label{fig:model}
\end{figure*}

\noindent The BlurRing method is based on using a random sampling of
the parameter space of the remaining parameters. The method is
introduced through the use of a three dimensional example. The example
model is built by adding a Gaussian signal to a uniform
background. The parameters of interest are the signal strength,
\textit{n}, the central value of the Gaussian, \textit{m}, and the
width of the Gaussian, \textit{w}. The full model is
\begin{equation*}
\label{eq:model}
n\cdot \Gauss(v,m,w)+(1-n) \cdot \Unif(v) , 
\end{equation*}
where $v$ is the fitted variable, $n \in [0,075]$ with a nominal value of 0.3, $m \in [-5,5]$ with
a nominal value of 0 and $w \in [0.1,4]$ with a nominal value of 4.
This model is used to generate 100 or 1000 events. The distribution of
events in the model is shown in figure \ref{fig:model}.

\begin{figure*}[!ht]
\noindent \begin{centering}
\includegraphics[width=.45\textwidth]{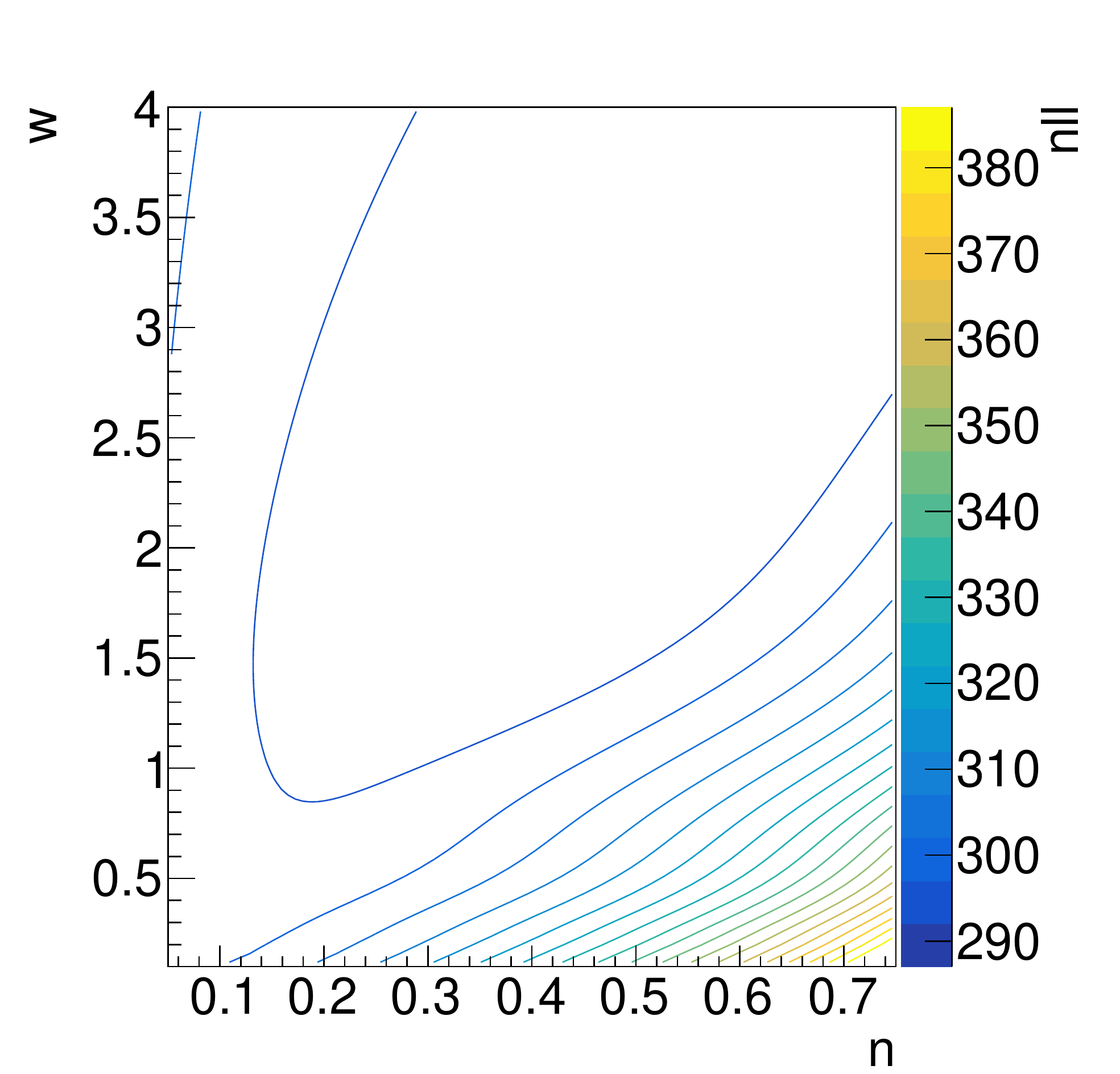}
\includegraphics[width=.45\textwidth]{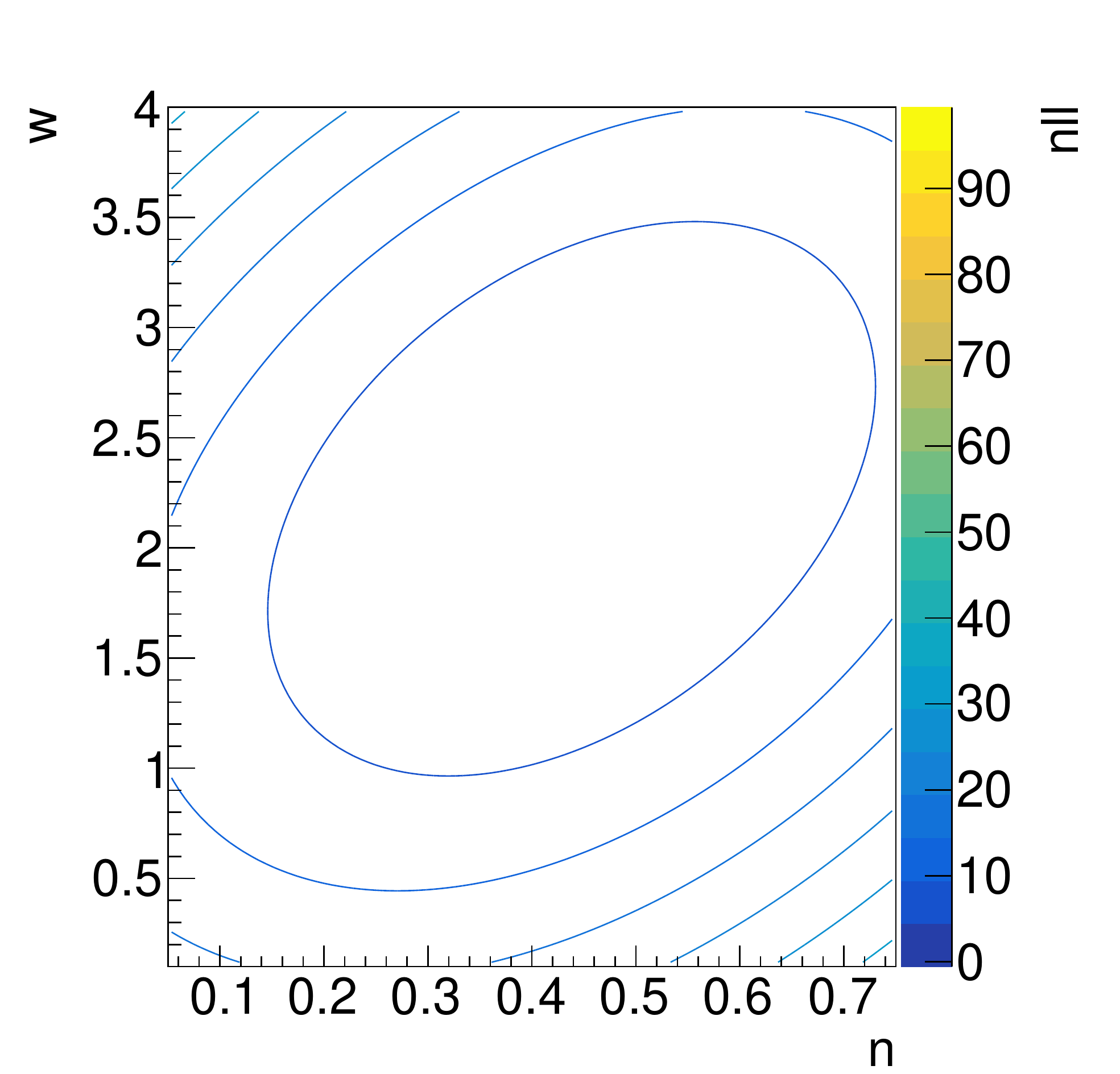}
\par\end{centering}
\caption{The Negative Log Likelihood (nll) contours for  $w$ vs $n$ (left) and its hessian approximation (right) for a model with 100 events.}
\label{fig:nllreg}
\end{figure*}

\noindent Figure \ref{fig:nllreg} shows the likelihood curves for a two dimensional slice of the parameter space;
\begin{equation*}
\label{eq:nll}
\mathcal{L}(n,w)=\frac{\mathcal{L}(m,w,m_{nom})}{\mathcal{L}(\hat{m},\hat{w},\hat{m})} .
\end{equation*}
In this case the third parameter is set to the nominal value of that
parameter. Although slices can be made in all three planes, the overall
structure of the likelihood in three dimensions is not clear from the
thee possible slices. The remaining slices can be found  in Appendix \ref{sec:appendix}. For a Gaussian likelihood, the shape of the
likelihood has a well known elliptical shape. At the minimum of any likelihood a Gaussian approximation can be constructed from the covariance matrix estimate. The Gaussian form of the likelihood created from the shape of the likelihood at the minimum of
the fit is called the Hessian approximation;
\begin{equation*}
\mathcal{L}_{\text{Hessian}} \propto \exp \left( \sum_{ij} \frac{1}{2} \cdot x_i\cdot H_{ij} \cdot x_j \right) ,
\end{equation*}
where $x$ is the vector of the sampled model parameters and $H_{ij}$ is the Hessian matrix 
\begin{equation*}
 H_{ij} = \left . \frac{\partial^2 ln \mathcal{L}(x)}{\partial x_i \partial x_j}\right \vert_{\hat{x}} .
\end{equation*}
 The Hessian approximation
of the likelihood for the example in this paper is shown in the same figure, and the elliptical shape is easily recognisable.

\section*{Sampling}
\noindent Instead of fixing the third parameter to the nominal
value or best fitted value, the BlurRing method uses a random sampling
method for the remaining parameters; 
\begin{equation*}
\label{eq:samplinggeneral}
\mathcal{L}(x|a,b),
\end{equation*}
where $x$ is the vector of the sampled model parameters and $a$ and $b$ are the scanned model parameters.
In the example of this paper,
there is only one remaining parameter which is sampled randomly, but
the method is independent of the number of sampled parameters.

\noindent In principle, the samples can be chosen arbitrarily. However, in order
for the resulting representation to be  indicative of the true shape
of the likelihood, the Likelihood itself is normalised and
then interpreted as a probability density function to draw the sample values
from. Samples are thus drawn from the Bayesian posterior density  function with a flat prior. This way, not only the shapes of the contours are reflective of
the Likelihood, but also the distribution of samples itself provides a
visual indication of the structure of the likelihood.

\noindent The technical implementation accompanying this paper is using a simple
rejection-sampling method, which is efficient enough for the example
provided here. Samples are drawn uniformly from the parameter space restricted to the $Z$-sigma hyper-ellipse to void outliers. 
Samples are accepted if a randomly chosen value is smaller
than the likelihood value at this point. For each accepted sample of
the Likelihood, the likelihood is evaluated on a grid of points, from
which the contours are then extracted.
For some $Z\sigma$ ellipse $U$ around the minimum $\hat{x}$, the
samples are chosen uniformly as $x\sim\Unif\left(U\right)$ and the
comparison value is chosen as
$y\sim\Unif\left(\left[1,Z\sigma\right]\right)$, such that
\begin{align*}
  P_{\textrm{accept}}\left(x\right) &= P\left(y<\frac{L\left(x\right)}{L\left(\hat{x}\right)}\right) = \frac{L\left(x\right)}{L\left(\hat{x}\right)}
\end{align*}
and hence
\begin{align*}
  y\sim\frac{L\left(x\right)}{\left(\hat{x}\right)}
\end{align*}
for the accepted points, where $x$ is the vector of the sampled model parameters.

\noindent A second implementation also accompanying this publication is using
a multivariate Gaussian as a Hessian approximation of the Likelihood
and is drawing samples from this approximation, which is possible
using standard random number generators without rejecting any points,
simply using
\begin{align*}
  y\sim\Gauss_L\left(x\right).
\end{align*}

\noindent A third method is also provided alongside in the
package. Employing the Gibbs-Sampling \cite{gibbs} method, Markov
Chain Monte Carlo (MCMC) sampling is used to more efficiently cover
the parameter space with samples. For each individual component, the
rejection sampler from the first implementation is used. 
Since the example discussed in this paper considers a single sampled parameter, Gibbs sampling is identical to rejection sampling for this case and is therefore not shown separately.

\begin{figure*}[!ht]
\noindent \begin{centering}
\includegraphics[width=0.45\textwidth]{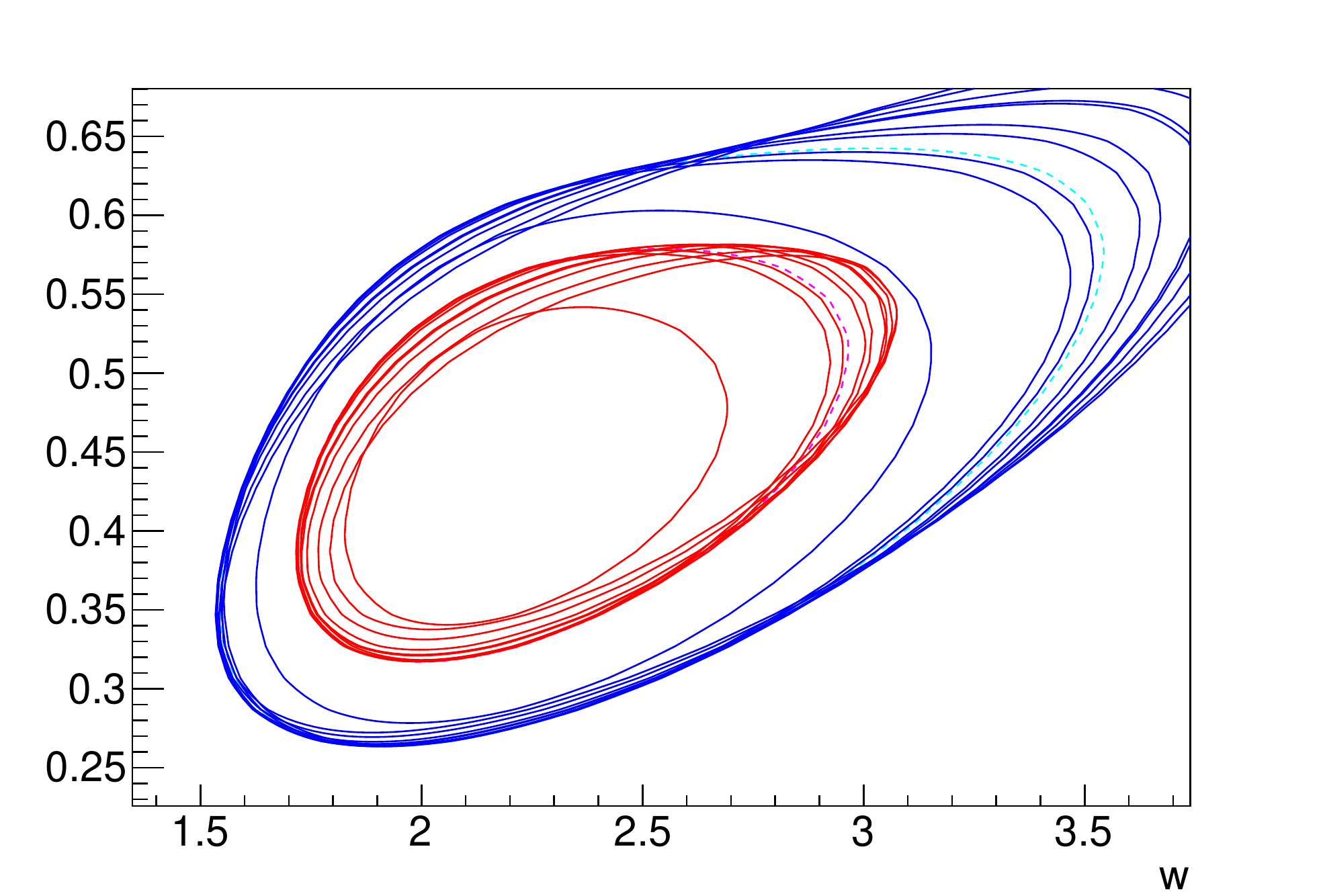}
\includegraphics[width=0.45\textwidth]{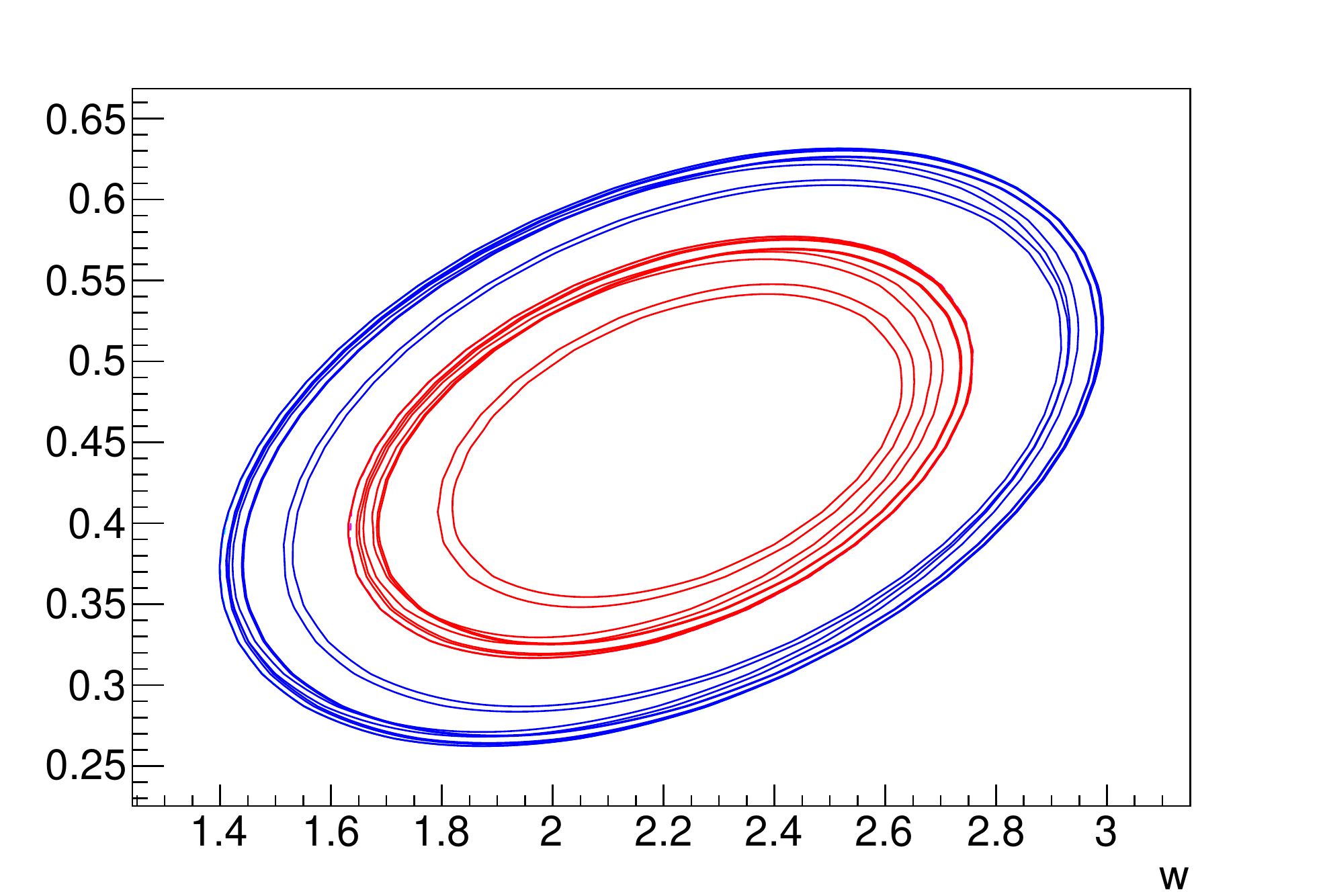}
\par\end{centering}
\caption{The Negative Log Likelihood (nll) contours with BlurRing for $w$ vs $n$ (left) and its hessian approximation (right) for a model with 100 events.}
\label{fig:blur}
\end{figure*}

\begin{figure*}[!ht]
\noindent \begin{centering}
\includegraphics[width=.45\textwidth]{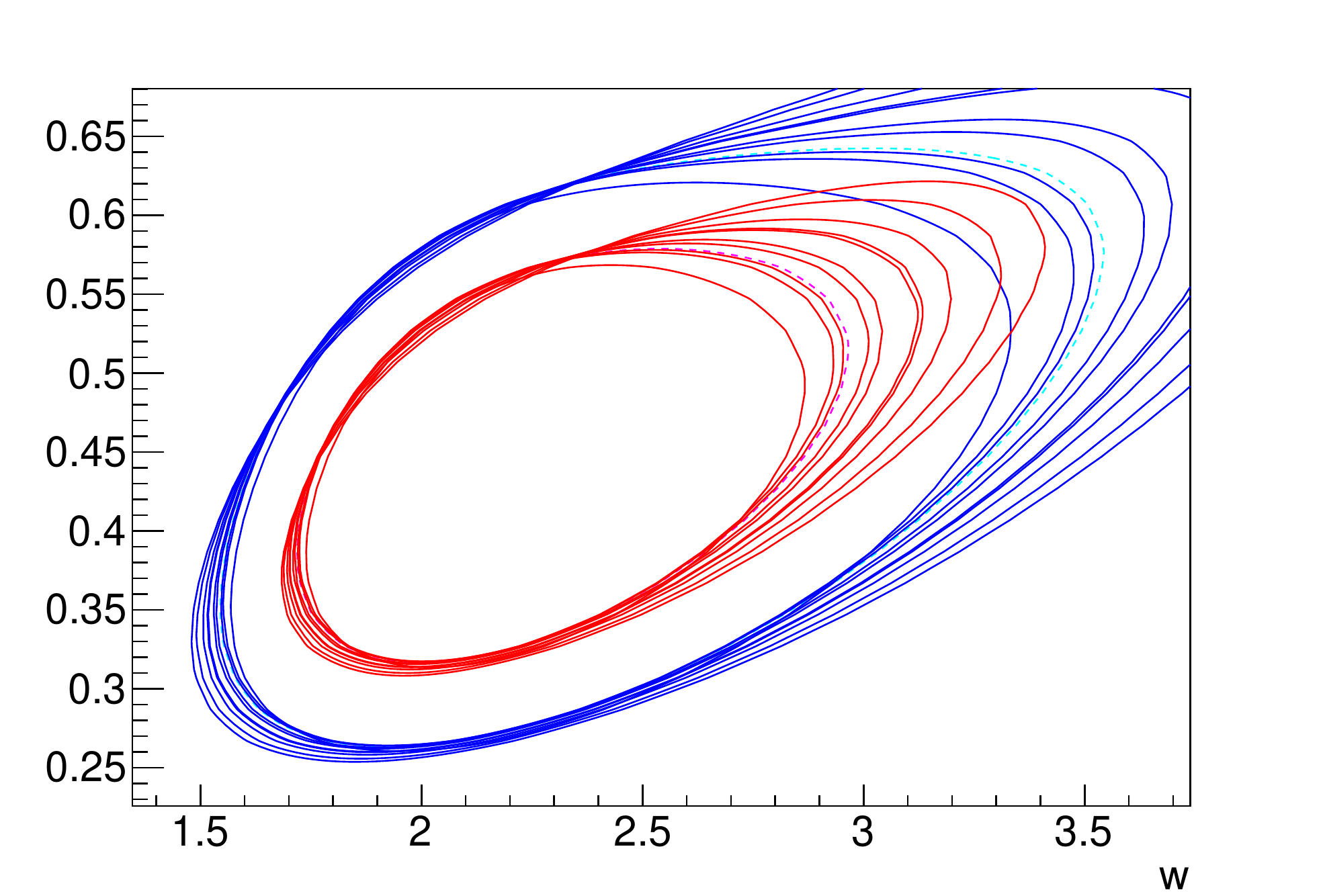}
\includegraphics[width=.45\textwidth]{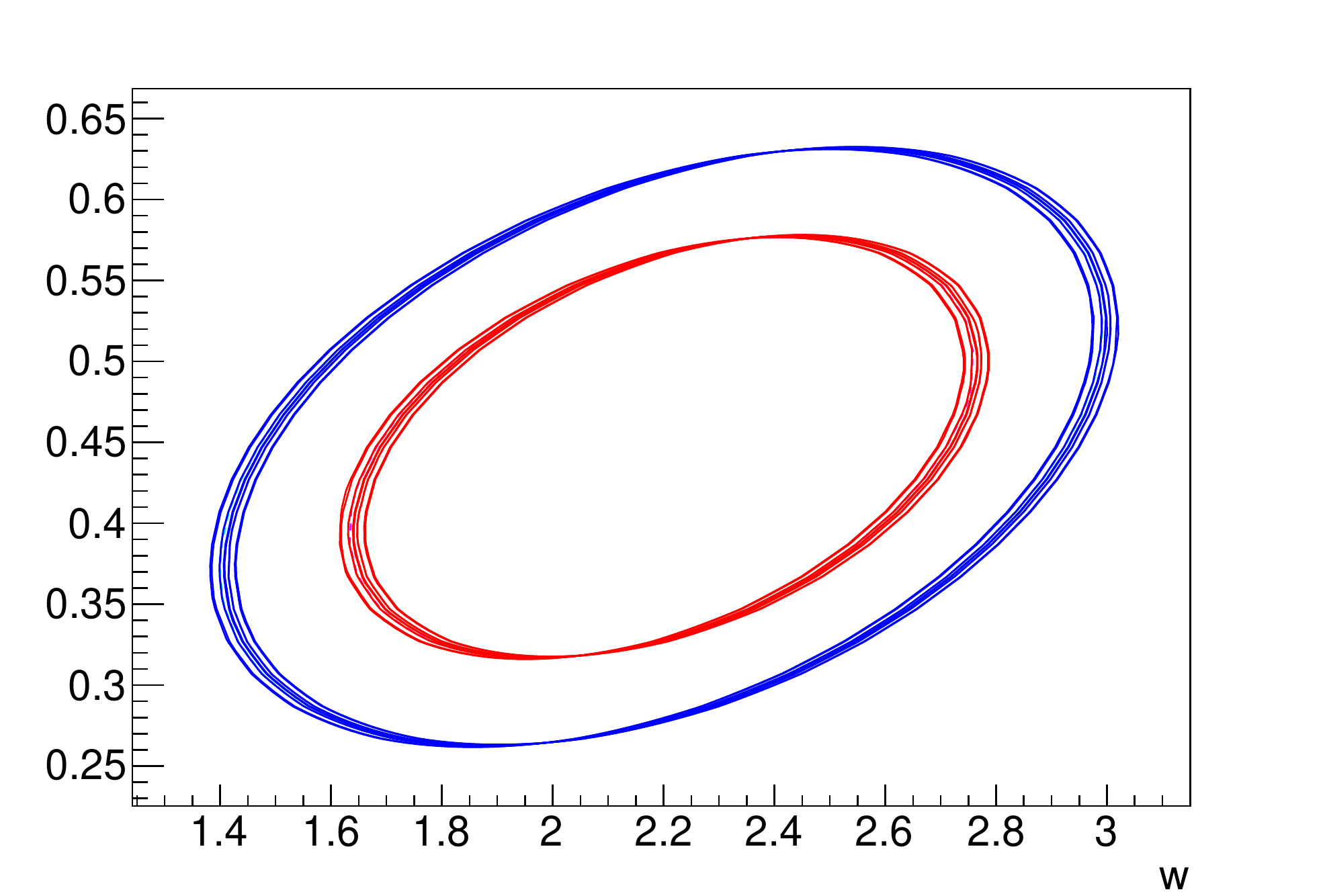}
\par\end{centering}
\caption{The Negative Log Likelihood (nll) contours normalised with BlurRing for $w$ vs $n$ (left) and its hessian approximation (right) for a model with 100 events.}
\label{fig:nornalisedblur}
\end{figure*}

\noindent In cases of few parameters and very complicated shapes,
rejection sampling is the safest and most accurate, but also slowest
option. For higher dimensional problems, Gibbs sampling should be
used, where the full likelihood structure is still preserved, but the
independence of the individual samples is not guaranteed. For problem
with an extremely large parameter space, the hessian sampling will
provide a very efficient method at the cost of some of the elegance of
the BlurRing method.

\section*{Visualisation}
\noindent By drawing the likelihood contours for randomly sampled
values of the remaining parameters, the 
effect of varying  profiled parameters 
becomes more clear. For the Hessian approximation case, where an elliptical
shape is expected, each slice of two parameters clearly shows the
expected shape as can be seen in figure \ref{fig:blur}. The
dotted lines give the one and two sigma contours of the likelihood
where the remaining parameters are fixed to their nominal
values. Figure \ref{fig:blur} also shows the one and two sigma
likelihood contours for the full likelihood, where the non-Gaussian
shape is very clear. In figure \ref{fig:nornalisedblur}  the likelihood contours are normalised to the minimum conditional of the sample values to give information purely on the distortion of the shape of the likelihood contour. Additional figures can be found in Appendix \ref{sec:appendix}.

Not only does this method represent the full
likelihood in a more comprehensive way, the spread in any direction
gives information related to the uncertainty on confidence interval
given on any single or couple of parameters. A large spread in the
likelihood curves in the BlurRing plots indicate that a small change
in the value of the not-plotted parameters can have a large effect on
the confidence interval determined from the likelihood curve.

Unlike scans where the remaining parameters are fixed to their nominal
values, the BlurRing method does not use a simplified model. While
profiled likelihoods extracted from simultaneous fits only show the
true likelihood curves near the minimum of the fit, while
incorporating the BlurRing method allows the likelihood curves to be
correctly presented throughout the full parameter space.

New information can be obtained for the plots about the stability of
the models, and correctness of the likelihood curves.

\section*{Conclusions}
\noindent The BlurRing method allows for multi-dimensional likelihood
visualisation from which addition information about the likelihood can
be extracted. The spread in any direction of the likelihood curves
gives information about the uncertainty on the confidence intervals
presented in the two-dimensional likelihood plots.

\section*{Acknowledgments}
We would like to thank Gottfried Herold and Dimitri Scheftelowitsch for providing useful hints on sampling implementations. We would also like to thank Glen Cowan and Wouter Verkerke for interesting discussions. Finally, we would like to thank the DESY and Nikhef institutes for giving us the opportunity to pursue this endeavour.  

\end{multicols}

\bibliography{main}

\clearpage
\appendix
\section{Appendix: Additional figures}
\label{sec:appendix}

\begin{figure*}[!ht]
\noindent \begin{centering}
\includegraphics[width=.3\textwidth]{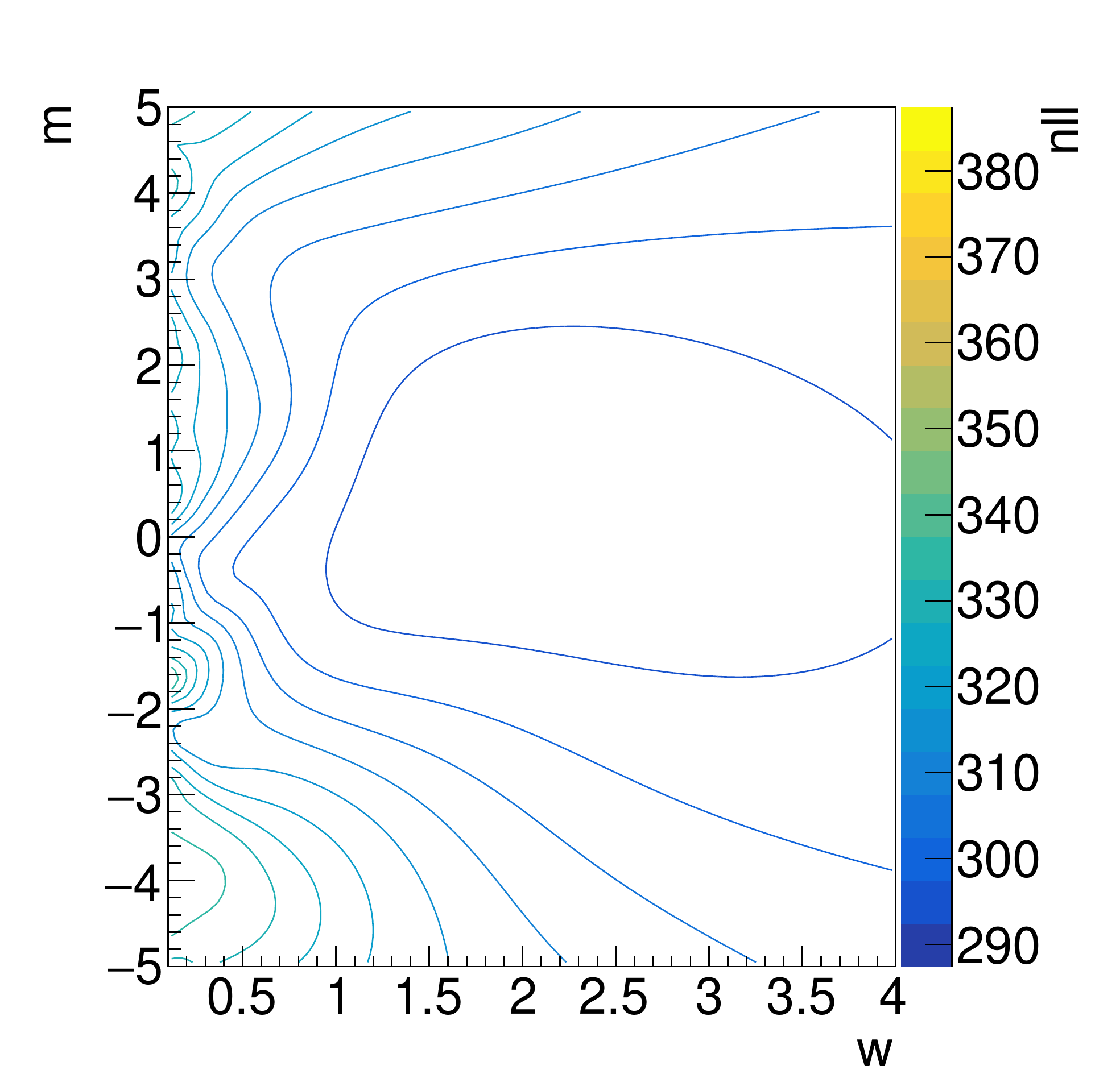}
\includegraphics[width=.3\textwidth]{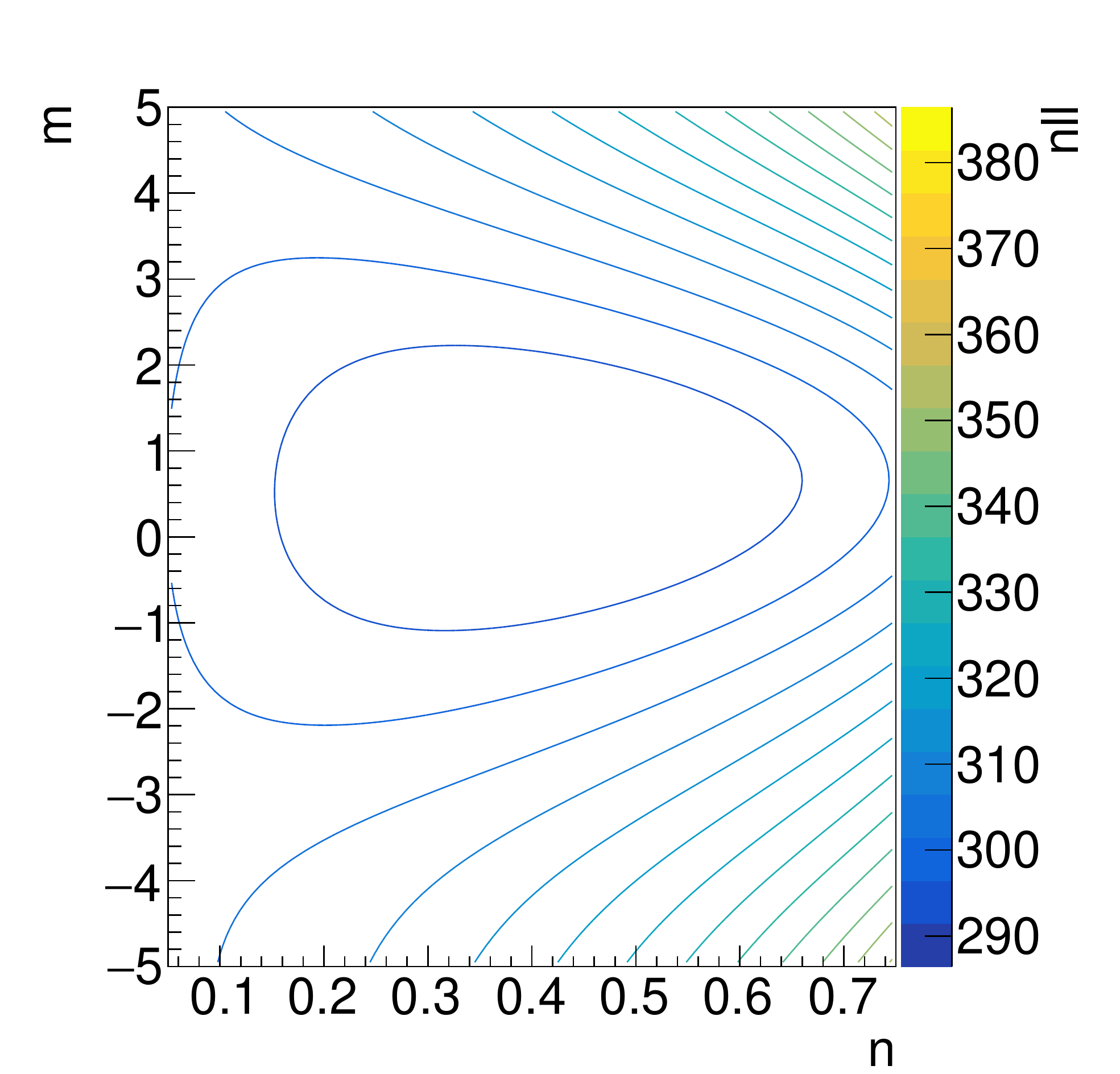}
\includegraphics[width=.3\textwidth]{newPlots100events/nll_scan_n_w.pdf}
\par\end{centering}
\caption{Negative Log Likelihood (nll) contours for $m$ vs $w$ (left) $m$ vs $n$ (middle) and $w$ vs $n$ (right) for a model with 100 events.}

\end{figure*}

\begin{figure*}[!ht]
\noindent \begin{centering}
\includegraphics[width=.3\textwidth]{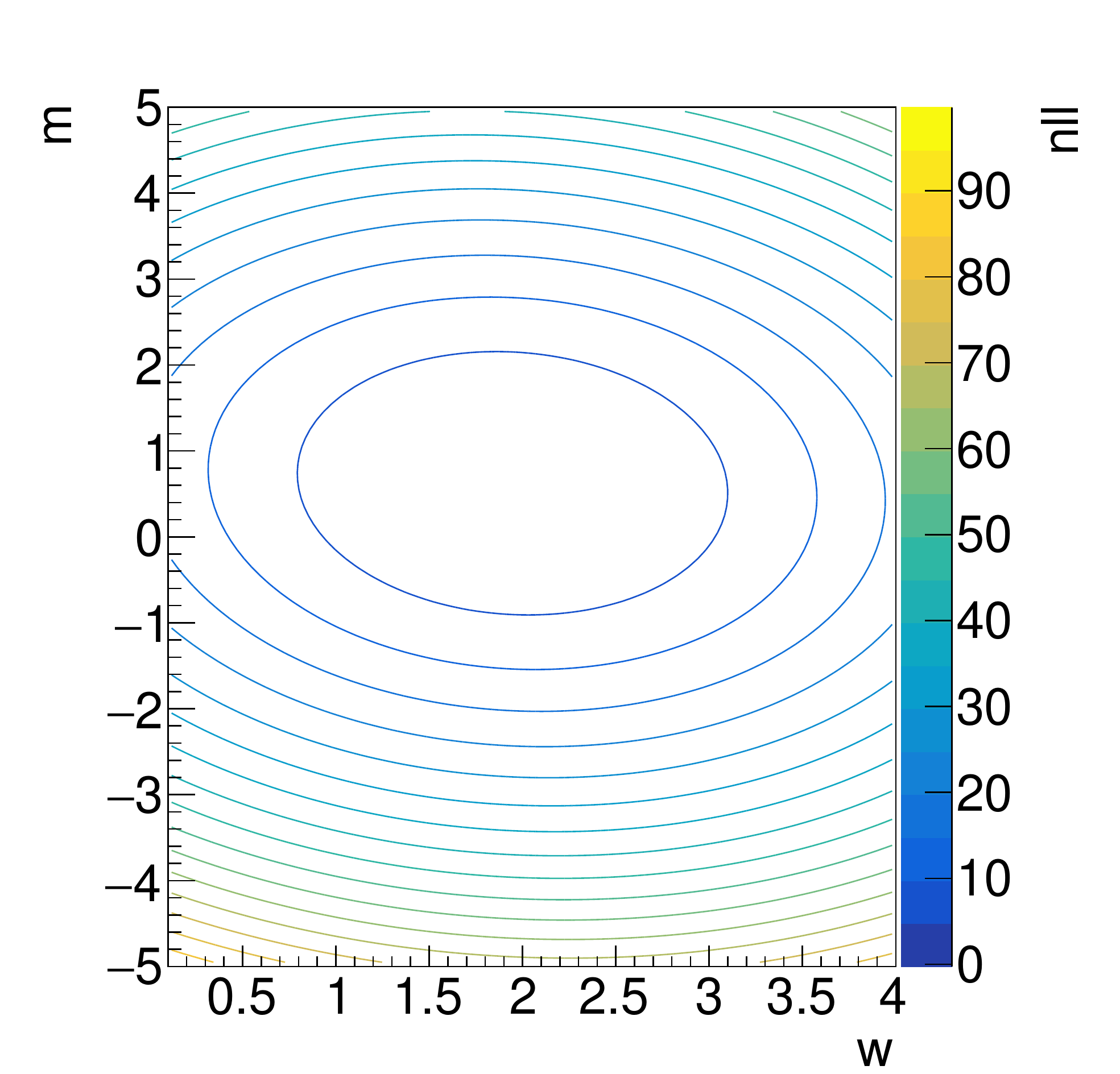}
\includegraphics[width=.3\textwidth]{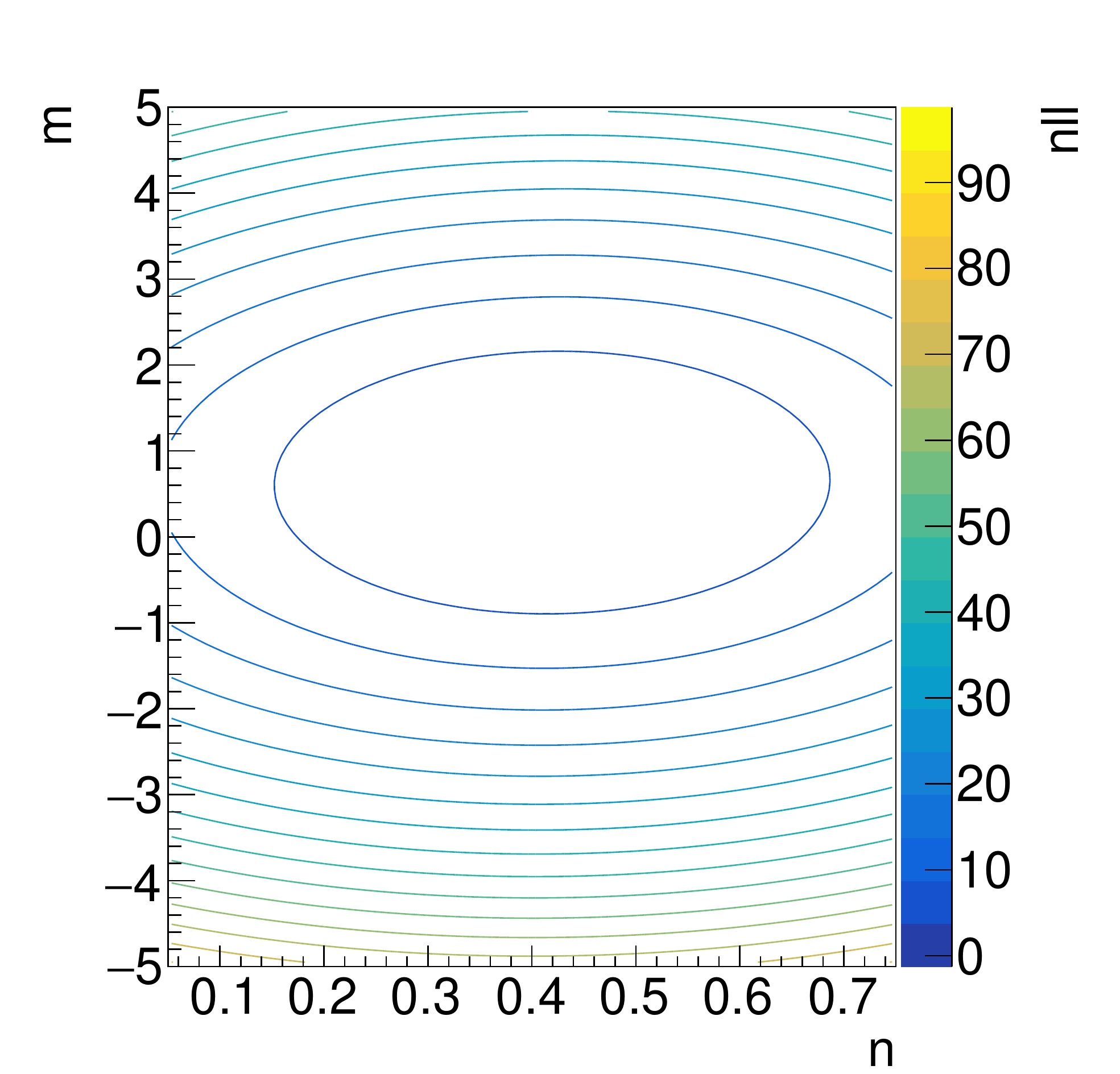}
\includegraphics[width=.3\textwidth]{newPlots100events/nll_hessian_scan_n_w.pdf}
\par\end{centering}
\caption{Hessian approximation of Negative Log Likelihood (nll) contours for $m$ vs $w$ (left) $m$ vs $n$ (middle) and $w$ vs $n$ (right) for a model with 100 events.}

\end{figure*}

\begin{figure*}[!ht]
\noindent \begin{centering}
\includegraphics[width=.45\textwidth]{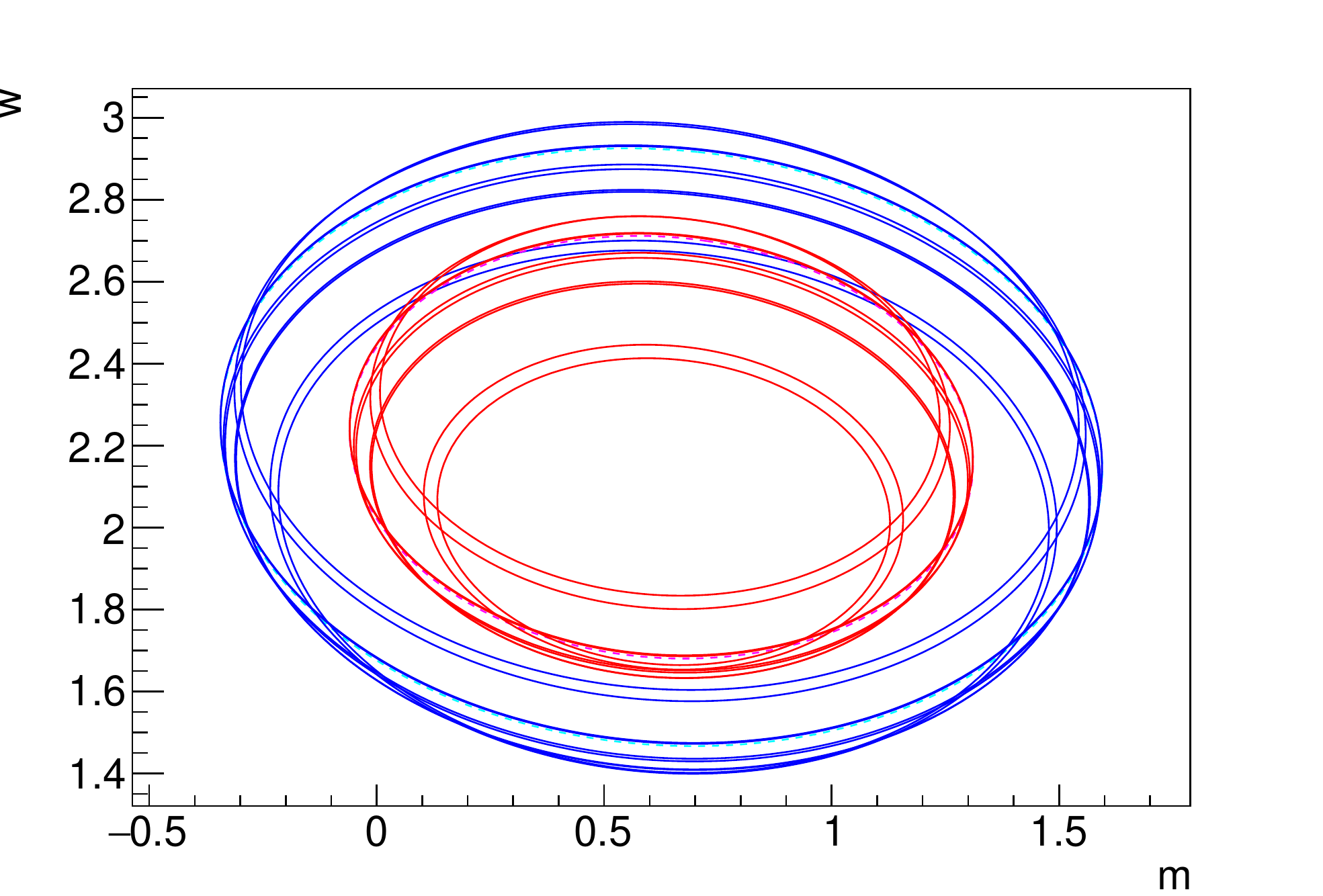}
\includegraphics[width=.45\textwidth]{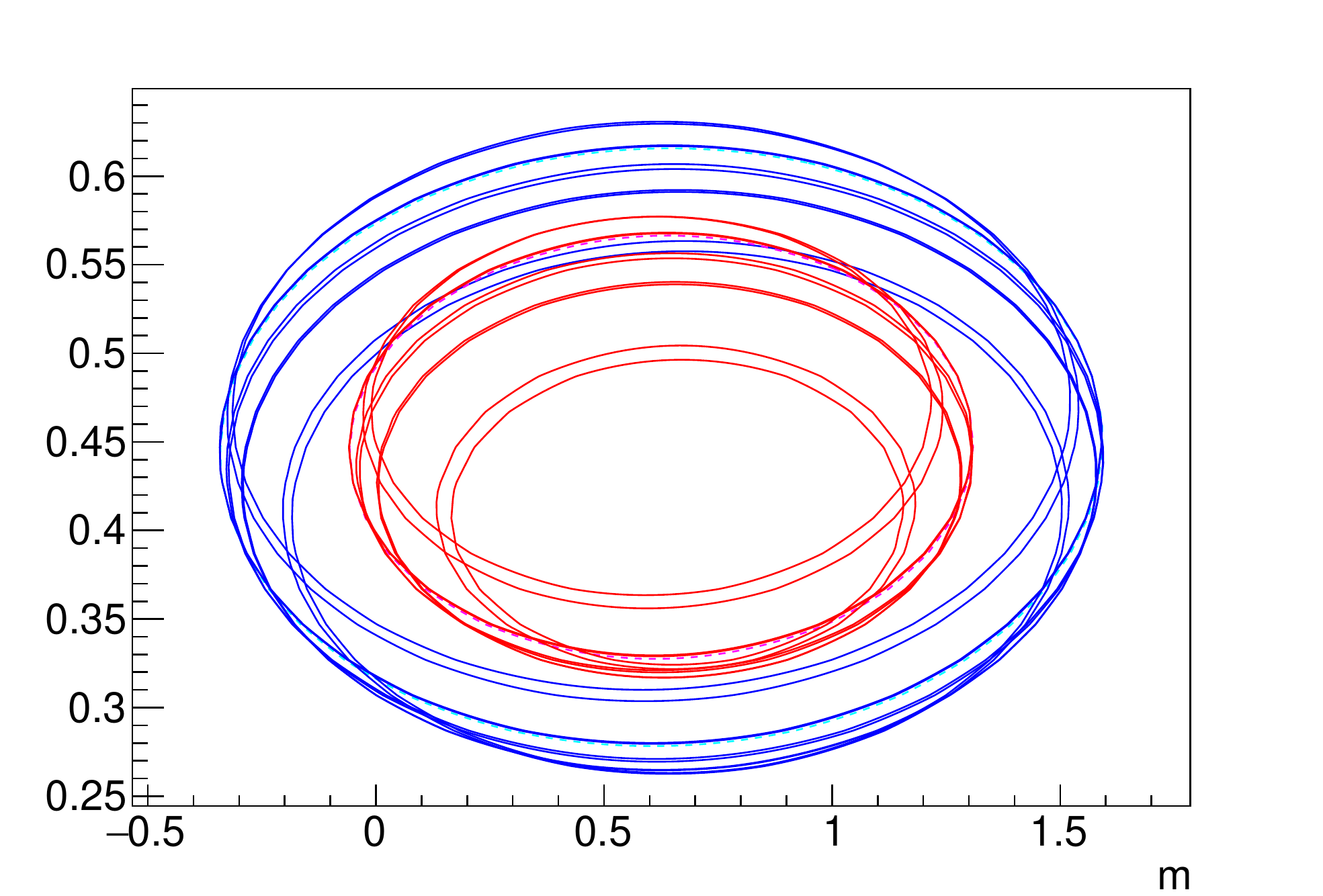}
\includegraphics[width=.45\textwidth]{newPlots100events/nll_model_hessian_hessian_data_w_n_absNorm.pdf}
\par\end{centering}
\caption{Hessian approximation of Negative Log Likelihood (nll) contours with BlurRing for $m$ vs $w$ (left) $m$ vs $n$ (right) and $w$ vs $n$ (bottom) for a model with 100 events.}
\end{figure*}

\begin{figure*}[!ht]
\noindent \begin{centering}
\includegraphics[width=.45\textwidth]{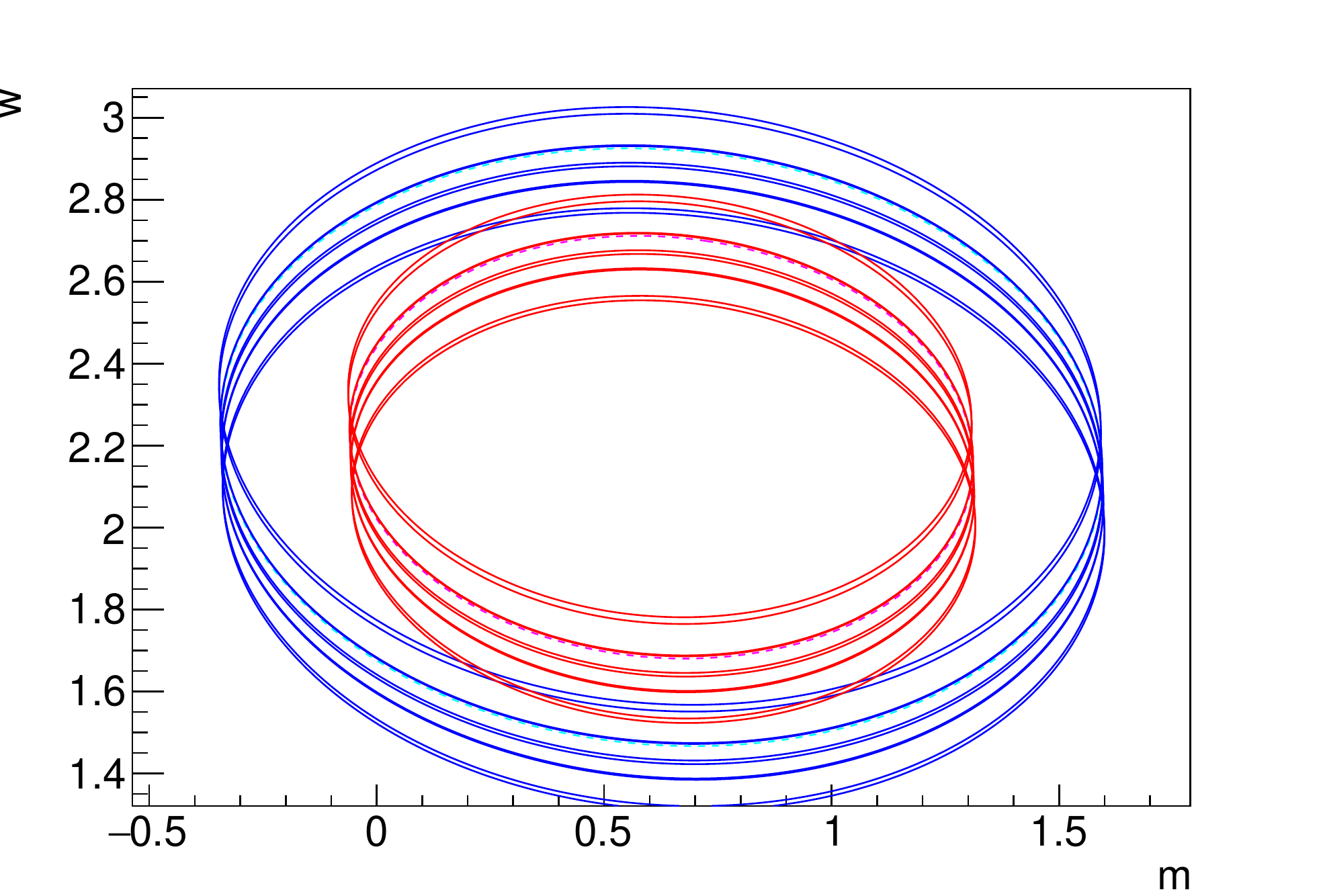}
\includegraphics[width=.45\textwidth]{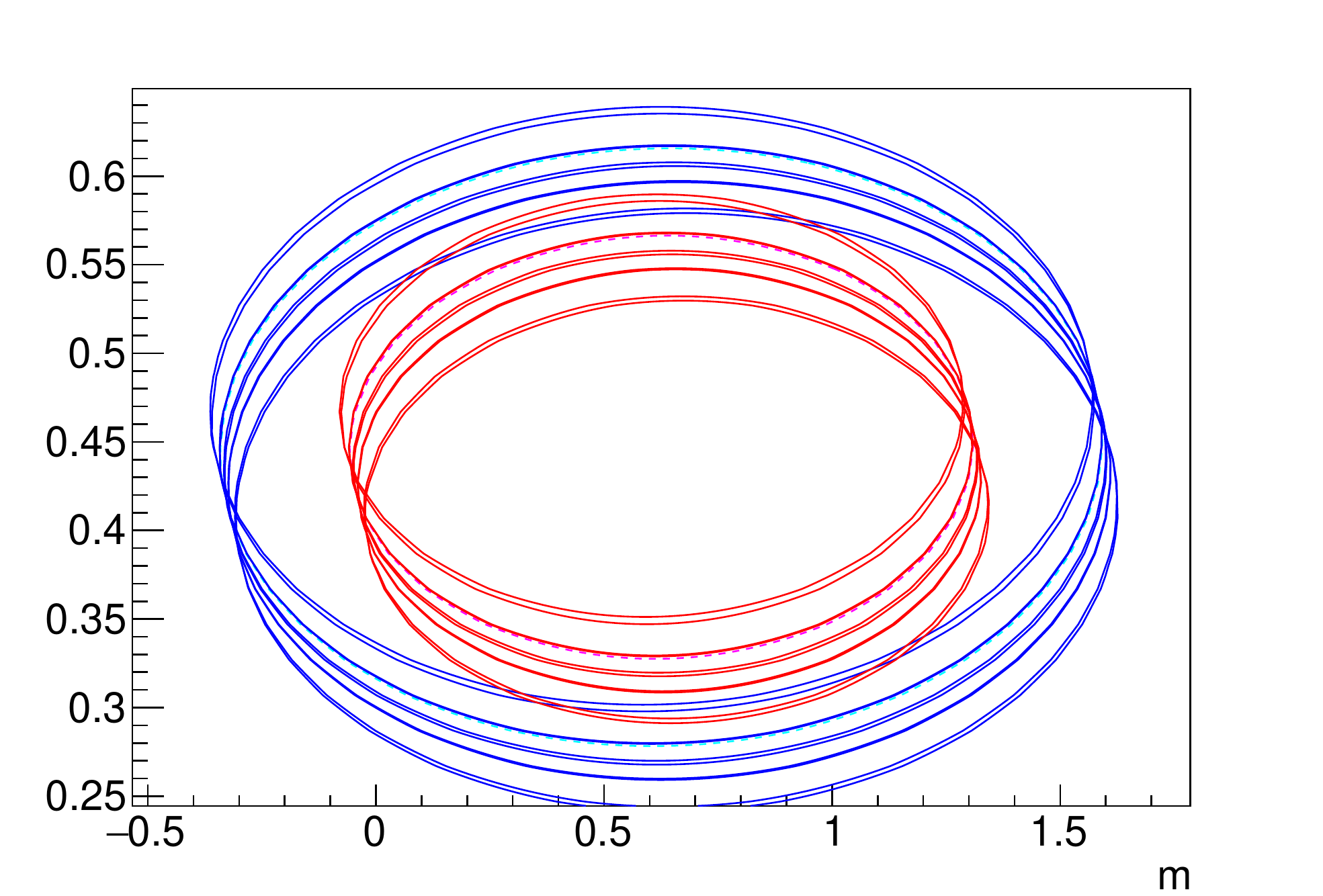}
\includegraphics[width=.45\textwidth]{newPlots100events/nll_model_hessian_hessian_data_w_n_relNorm.pdf}
\par\end{centering}
\caption{Hessian approximation of Negative Log Likelihood (nll) contours normalised with BlurRing for $m$ vs $w$ (left) $m$ vs $n$ (right) and $w$ vs $n$ (bottom) for a model with 100 events.}
\end{figure*}

\begin{figure*}[!ht]
\noindent \begin{centering}
\includegraphics[width=.45\textwidth]{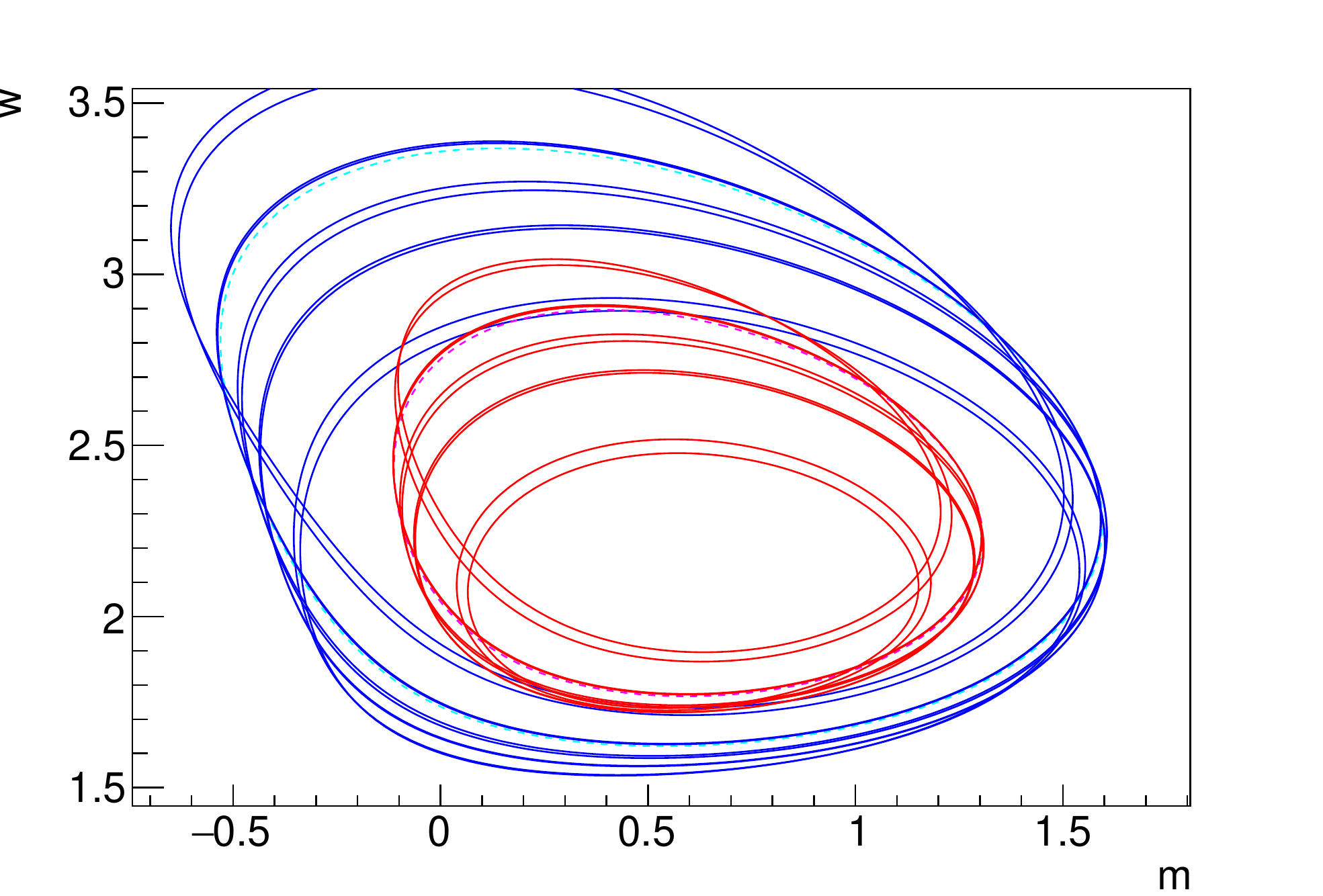}
\includegraphics[width=.45\textwidth]{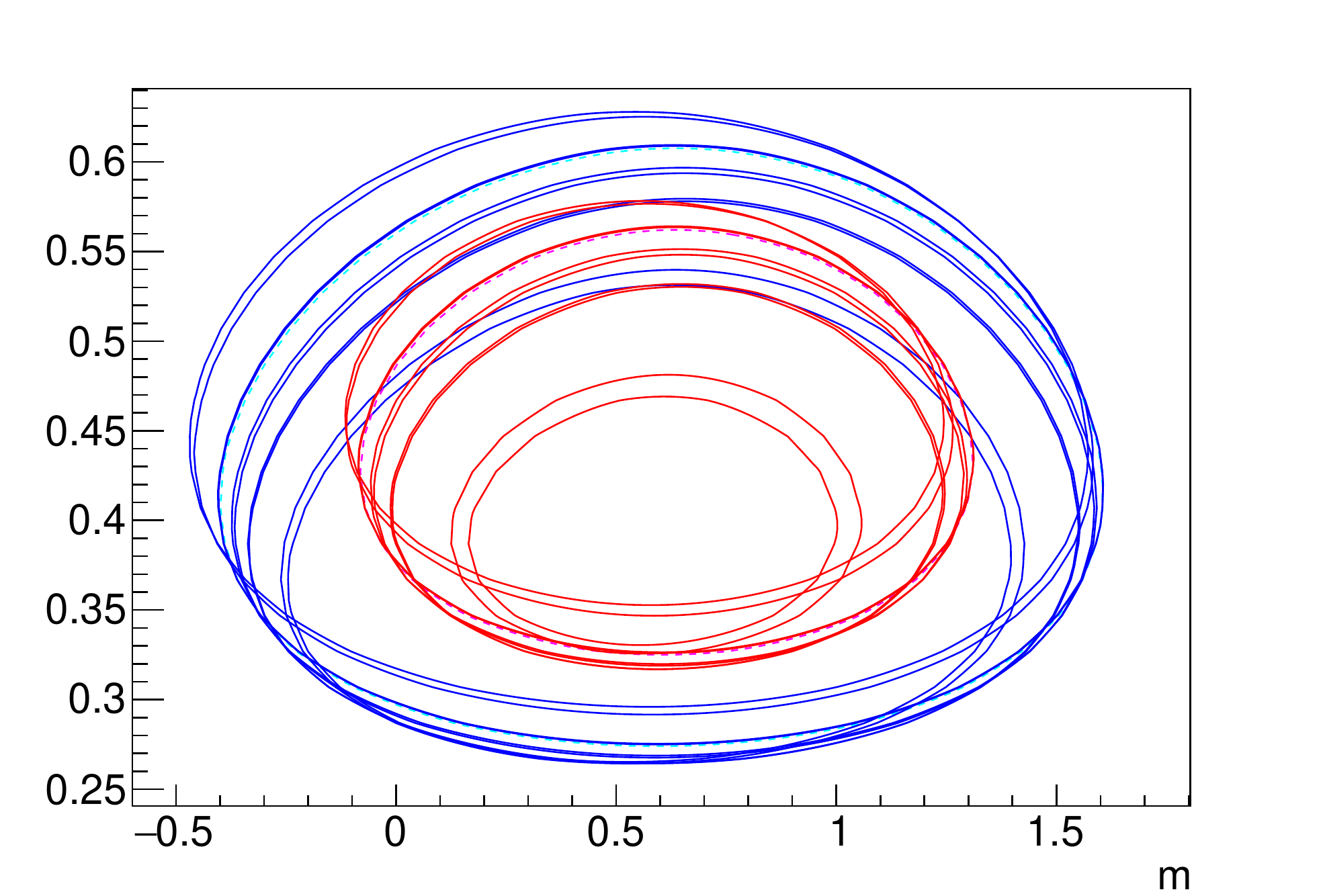}
\includegraphics[width=.45\textwidth]{newPlots100events/nll_model_modelData_w_n_absNorm.pdf}
\par\end{centering}
\caption{Negative Log Likelihood (nll) contours with BlurRing for $m$ vs $w$ (left) $m$ vs $n$ (right) and $w$ vs $n$ (bottom) for a model with 100 events.}
\end{figure*}

\begin{figure*}[!ht]
\noindent \begin{centering}
\includegraphics[width=.45\textwidth]{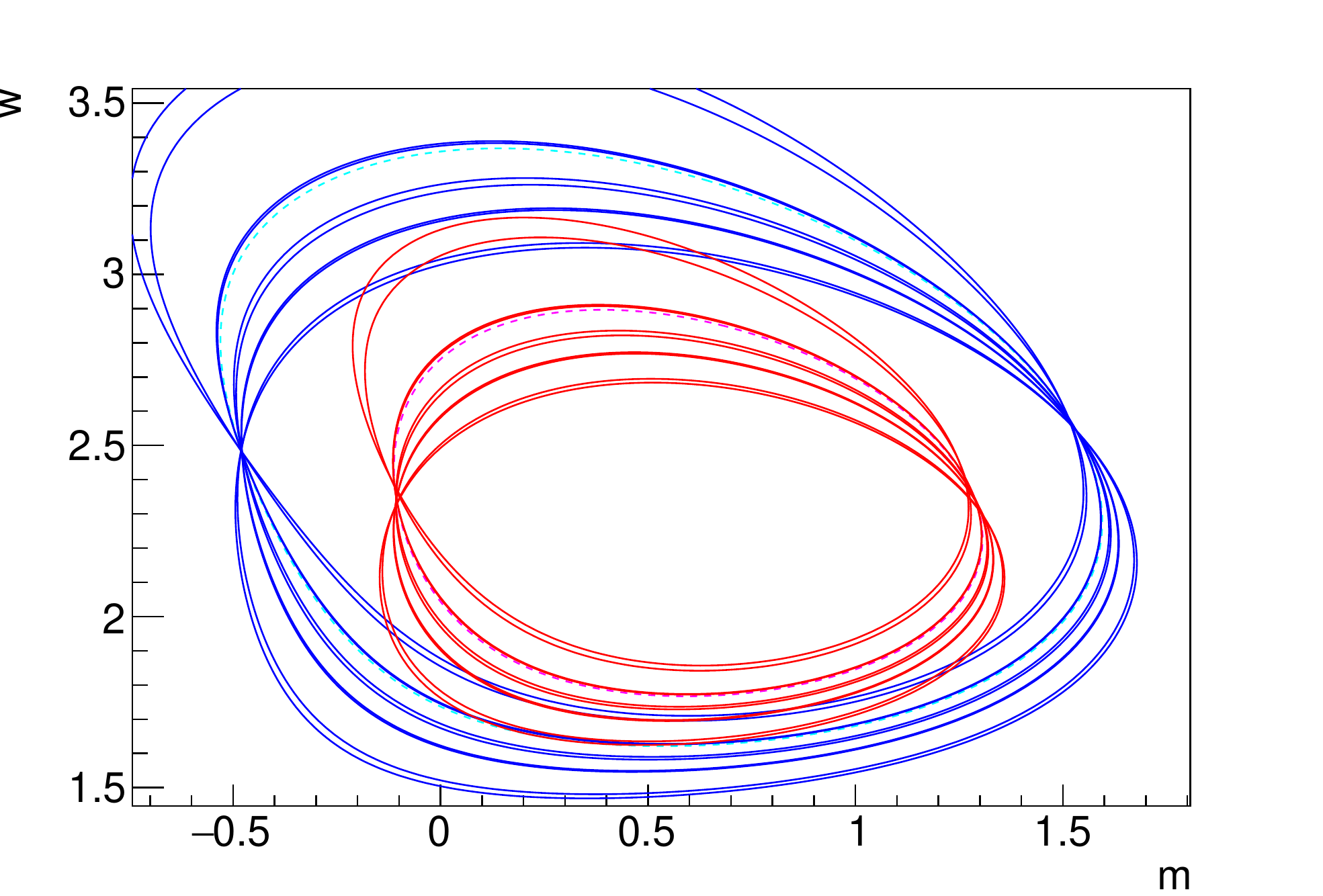}
\includegraphics[width=.45\textwidth]{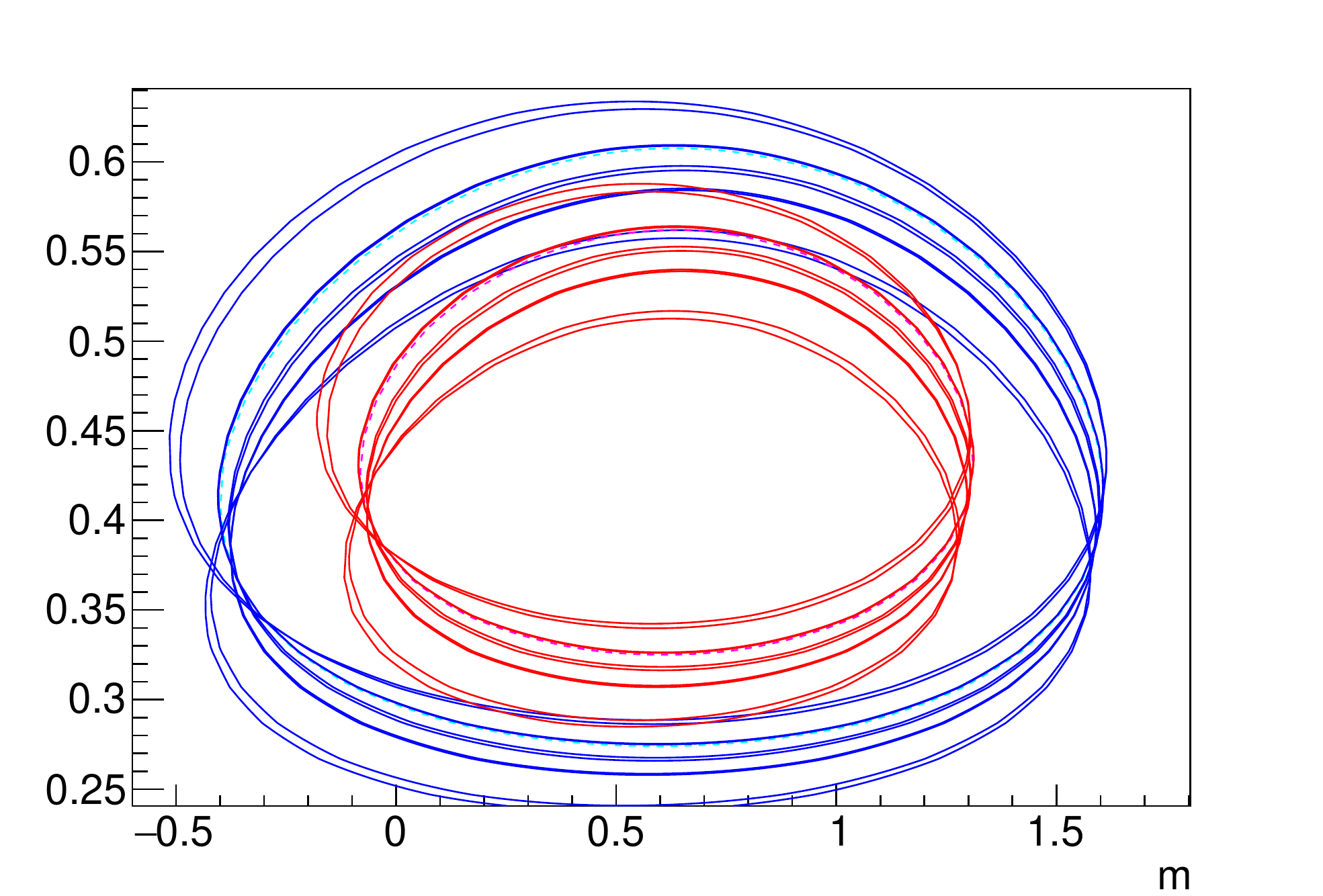}
\includegraphics[width=.45\textwidth]{newPlots100events/nll_model_modelData_w_n_relNorm.pdf}
\par\end{centering}
\caption{Negative Log Likelihood (nll) contours normalised with BlurRing for $m$ vs $w$ (left) $m$ vs $n$ (right) and $w$ vs $n$ (bottom) for a model with 100 events.}
\end{figure*}

\end{document}